\documentclass[a4paper,11pt]{article}
\usepackage{pos}
\usepackage{mathtools}
\usepackage{textgreek}

\title{CMS Tracker Alignment: Legacy results from LHC Run 2 and Run 3 prospects}
 \ShortTitle{CMS Tracker Alignment: Legacy results from LHC Run 2 and Run 3 prospects}

\author*[a,1]{Sandra Consuegra Rodríguez}

\affiliation[a]{Deutsches Elektronen-Synchrotron,\\
  Notkestraße 85, 22607 Hamburg, Germany}
  
\note{On behalf of the CMS Collaboration}  

\emailAdd{sandra.consuegra.rodriguez@desy.de}

\abstract{The inner tracking system of the CMS experiment, which comprises Silicon Pixel and Silicon Strip detectors, is designed to provide a precise measurement of the momentum of charged particles and to reconstruct the primary and secondary vertices. The movements of the different substructures of the tracker detectors driven by the operating conditions during data taking, require to regularly update the detector geometry in order to accurately describe the position, orientation, and curvature of the tracker modules.
\\
The procedure in which new parameters of the tracker geometry are determined is known as alignment of the tracker. The alignment procedure is performed several times during data taking using reconstructed tracks from collisions and cosmic rays data, and later on, further refined after the data taking period is finished.  The tracker alignment performance corresponding to the ultimate accuracy of the alignment calibration for the legacy reprocessing of the CMS Run 2 data is presented.  The data-driven methods used to derive the alignment parameters and the set of validations that monitor the performance of physics observables after the alignment are reviewed. Finally, the prospects for the alignment calibration during the upcoming run of the LHC, where more challenging operation conditions are expected,  will be addressed.
}

\FullConference{%
The European Physical Society Conference on High Energy Physics (EPS-HEP2021)\\
26-30 July 2021\\

Hamburg, Germany (virtual meeting)
}

\begin{document}
\maketitle

\section{The CMS tracker} 
The CMS Tracker comprises both the silicon pixel and strip detectors within a length of 5.8 m and a diameter of 2.5 m \cite{citation1}.  The silicon pixel detector is the closest detector in proximity to the interaction point and consists of a barrel region (BPIX) and two forward endcaps (FPIX).  In the configuration operating until the end of the 2016 data-taking period (Phase-0 pixel detector),  the BPIX consisted of three layers and the FPIX of two disks in each of the two endcaps,  for a total of 1440 modules.  The upgraded pixel detector in operation since 2017 (Phase-1 pixel detector) has an additional barrel layer in BPIX and an additional disk on each of the two endcaps of the FPIX, which results in a total of 1856 modules  \cite{citation2}. The silicon strip detector consists of 15 148 modules and is composed of four subsystems: the Tracker Inner Barrel and Disks (TIB and TID), the Tracker Outer Barrel (TOB), and the Tracker EndCaps (TEC). \\
The determination of the trajectory of charged particles (tracks) from signals (hits) in the tracker,  referred to as tracking,  is instrumental for the process of event reconstruction in CMS, with a good tracking performance of particular relevance for many physics analyses. The measurement of the momentum of the tracks relies on an accurate determination of the track curvature induced by the magnetic field, which in turn requires a carefully calibrated detector.  The resolution of the momentum measurement is affected by multiple scattering and by the limited knowledge of the position of each of the track hits.  The latter has contributions from two main components, the spatial resolution of the modules, known as intrinsic hit resolution ($\approx$ 10 to 30 μm), and the precision that comes from the limited knowledge of the position and orientation of the modules,  known as alignment position errors.  The reduction of this last component well below the intrinsic hit resolution is needed in order to fully exploit the remarkable resolution of the CMS silicon modules.  Starting from a precision after installation of  $\mathscr{O}$(100 μm),  this is accomplished through a track-based alignment.

\section{Alignment of CMS tracker}
The alignment of the CMS tracker using reconstructed charged particles,  known as track-based alignment,
constitutes a major challenge due to the enormous number of degrees of freedom involved.  
A measured hit position is assigned to every hit registered in the detector and a set of tracks is formed from the combination of these hits.  To each of these tracks, a set of track parameters (e.g.,  parameters related to the track curvature and deflection angles due to multiple scattering) is associated. The fitted tracks depend on so-called alignment or global parameters as well, with nine of these parameters needed per sensor to set the coordinates of the sensor center,  rotational angles,  and surface deformations  \cite{citation3}. The track-based alignment method follows a least-square approach, minimizing the sum of squares of the normalized track-hit residuals from a set of tracks.  The track hit-residuals are obtained by subtracting the projection of the track prediction which depends on the track and alignment parameters from the measured hit position.  The resulting system of equations for the alignment parameters obtained from the minimization of the track-hit residuals is solved through a global fit \cite{citation4}.  The tracks are then re-fitted assuming the new geometry defined by the updated set of alignment parameters and the measurement of the track momentum is also updated.  The validation of the new alignment conditions follows.

\section{Tracker alignment strategy for data}
The hierarchical structure of the CMS tracker allows aligning high-level mechanical structures up to single modules.  Each year,  before the beginning of the data taking for physics analysis,  at least the high-level structures of the tracker are aligned using the available cosmic ray data from the commissioning of the detector.  This initial alignment allows a first non-refined determination of the alignment constants,  which at this point can show large movements with respect to the initially assumed geometry,  due to temperature and magnetic field changes, as well as the reinstallation of detector components during the detector shutdown.  These alignment constants are then used online for the express processing of the data.  An automated alignment continuously monitors the high-level structure movements of the pixel detector and automatically corrects the geometry if the alignment corrections exceed certain thresholds.  A track-based alignment at a higher granularity level is also periodically run offline.  A heterogeneous sample of tracks crossing the detector at different angles and covering their full active area to relate different detector components is used.  The automated alignment is then refined with periodic updates from these campaigns happening in parallel.  After the data-taking period is finished,  the full statistics of the dataset collected during the year is used to provide the alignment conditions for the reprocessing of the data.  The ultimate accuracy of the alignment calibration is derived and used for the final or legacy reprocessing of the data.  For the 2016-2018 period,  up to $\approx$ 700k parameters and 220 geometries for different intervals of validity were determined,  covering the significant changes of the alignment conditions over time.

\section{Legacy results}
The tracker geometry obtained from each alignment fit can be compared to the starting geometry for the particular fit,  to identify unusual movements or systematic distortions artificially introduced by the $\chi^2$ minimization of the track-hit residuals.  Further validations are carried out, inspecting observables that are particularly sensitive to the quality of the alignment.  The alignment constants for the modeling of the data in simulation were derived separately for each data-taking year and tuned to emulate the effects of the residual misalignment left in data after the legacy reprocessing.  The comparison of the performance of the different sets of alignment constants used during data taking,  for the end-of-year reconstruction,  and the legacy reprocessing is illustrated in this section.  

\subsection{Tracking performance}
\begin{figure}[!ht]
    \centering
       \includegraphics[width=.49\textwidth]{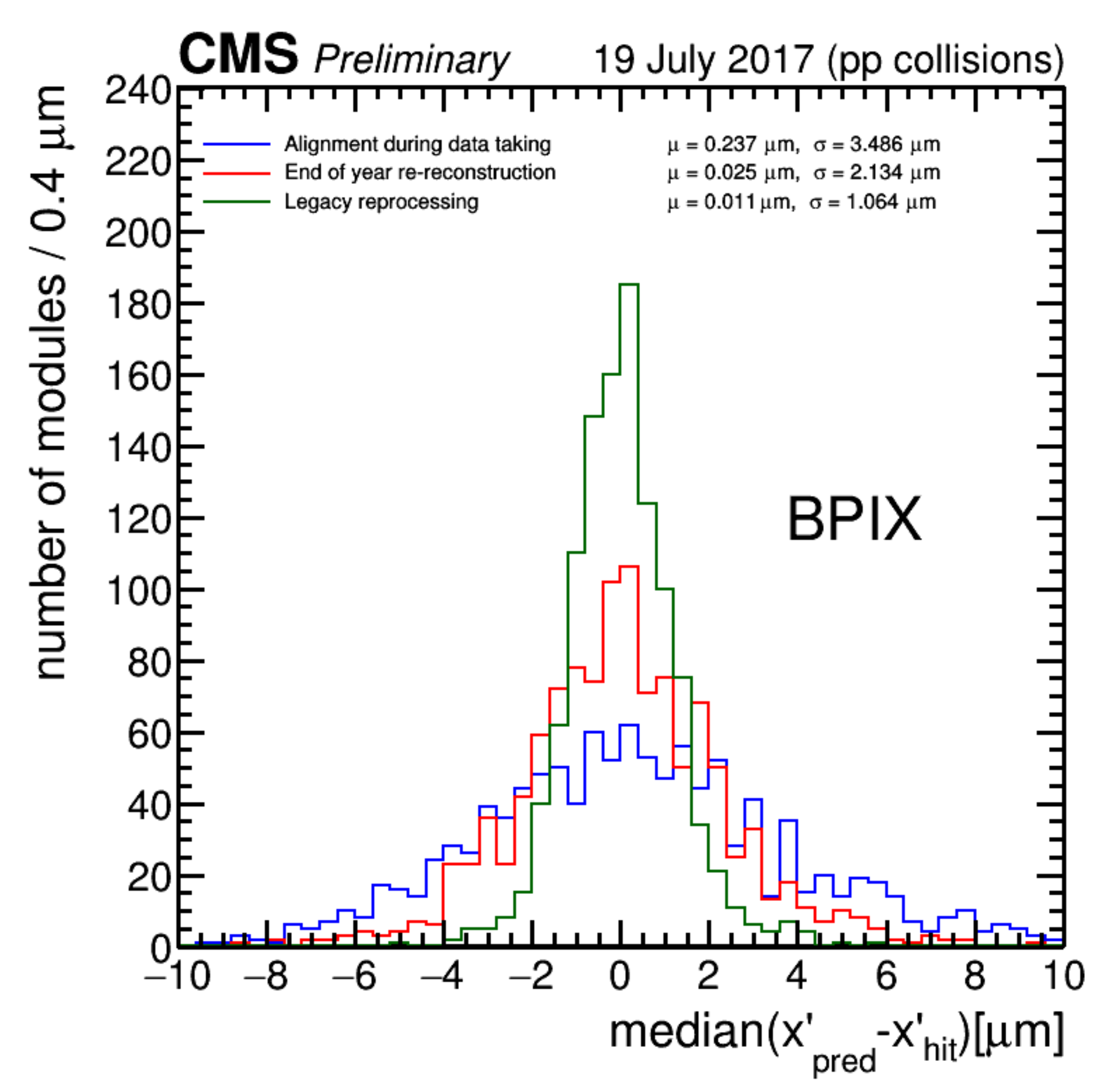}
         \includegraphics[width=.49\textwidth]{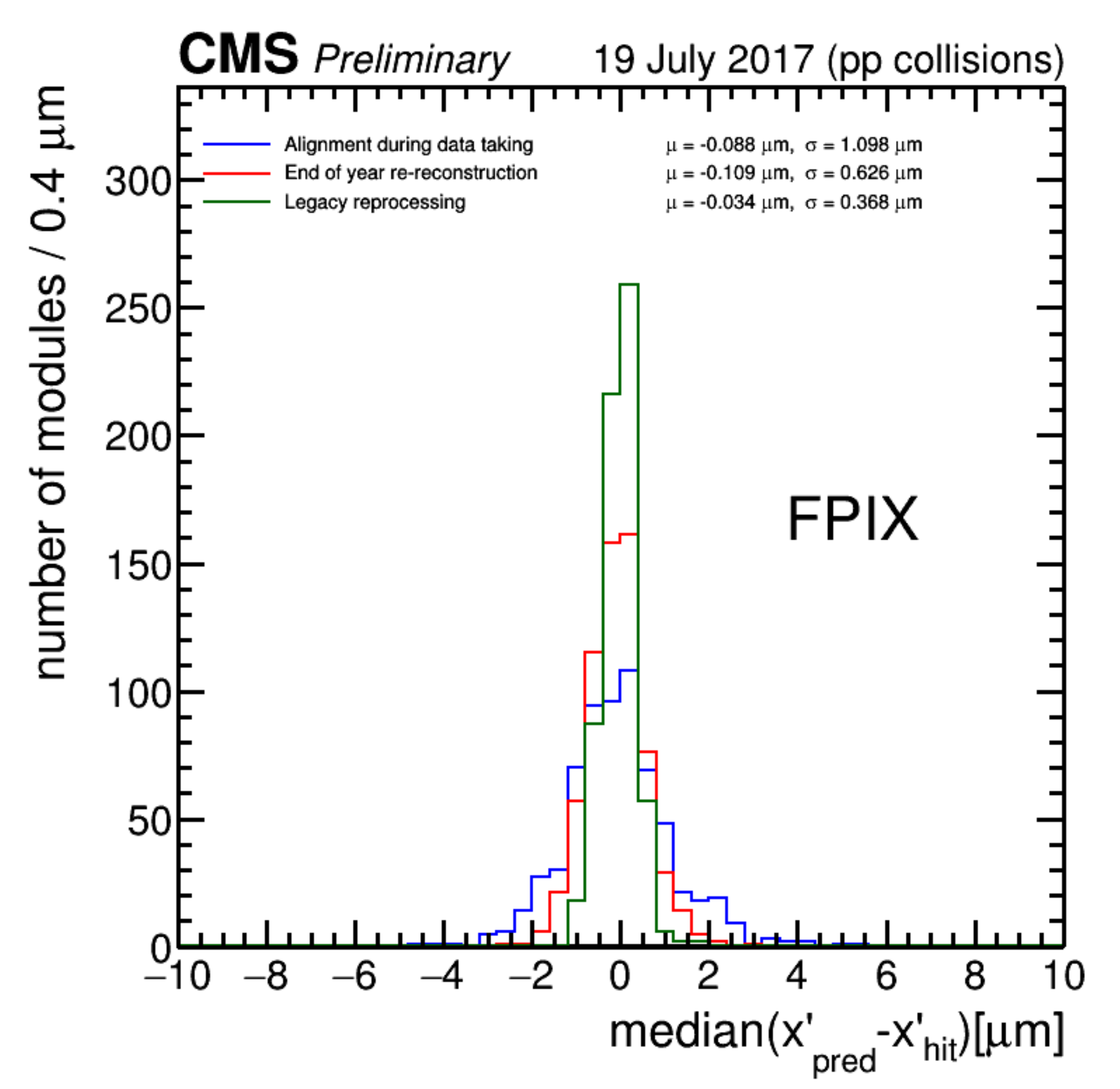}
       \caption{The distribution of median residuals is shown for the local-x coordinate in the barrel pixel (left) and forward pixel (right).  The alignment constants used for the  Run 2 legacy reprocessing (green) are compared with the ones used during data taking (blue) and for the end-of-year reconstruction (red).  The quoted means $\mu$ and standard deviations $\sigma$ correspond to the parameters of a Gaussian fit to the distributions \cite{citation5}.}
       \label{fig1}
     \end{figure}

The distribution of the median of the track-hit residuals per module (DMRs) constitutes a measure of the tracking performance.  Each track is refitted without the hit under consideration. Therefore, the computation of the residuals is said to be unbiased.   The distribution of the median should result in a narrow and centered distribution around zero since the track-based alignment performs the minimization of the sum of the track-hit residuals. The width of the distribution constitutes a measure of the local precision of the alignment calibration, while deviations of the mean from zero indicate possible biases due to change of conditions in the detector. The DMRs are monitored for all the tracker substructures,  as shown for the barrel and forward pixel detector in Figure \ref{fig1},  for the local x-coordinate of the modules.  A significant improvement for the legacy reprocessing over the alignment during data taking and the end-of-year reconstruction is observed.

\subsection{Vertexing performance}
\begin{figure}[!ht]
    \centering
       \includegraphics[width=1\textwidth]{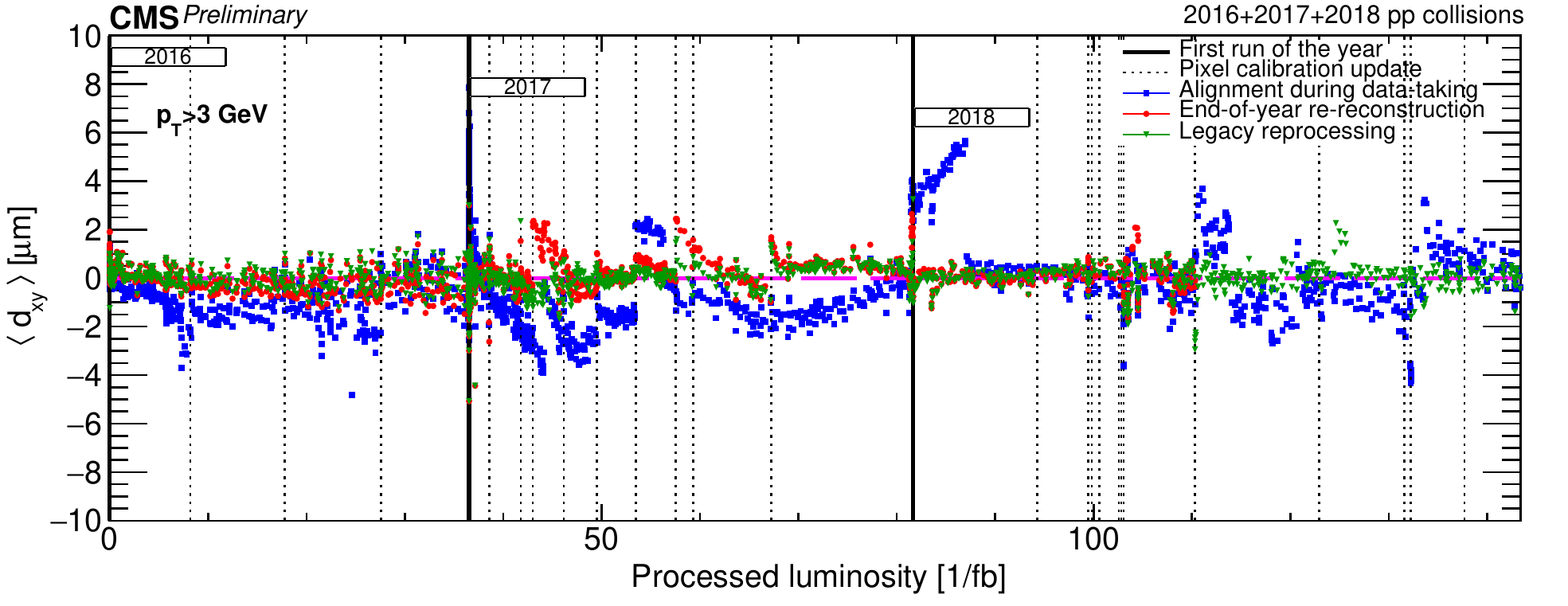}
       \caption{ The mean track-vertex impact parameter in the transverse plane $d_{xy}$ is shown as a function of the processed luminosity.   The vertical black lines indicate the first processed run for each data-taking year, and the vertical dotted lines indicate a change in the pixel tracker calibration \cite{citation6}.    
       }
       \label{fig2}
     \end{figure}

The effect of the alignment calibration on the reconstruction of physics objects is also studied. The distance between tracks and the vertex reconstructed without the track under scrutiny (unbiased track-vertex residuals) is studied to evaluate the performance of the alignment in the pixel detector, searching for potential biases in the primary vertex reconstruction. The mean value of the unbiased track-vertex residuals is shown in Figure \ref{fig2} for the transverse plane as a function of the processed luminosity.  The large alignment corrections that were needed at the beginning of each year of data taking can be observed.

\subsection{Monitoring of systematic distortions}
\begin{figure}[!ht]
    \centering
       \includegraphics[width=0.4\textwidth]{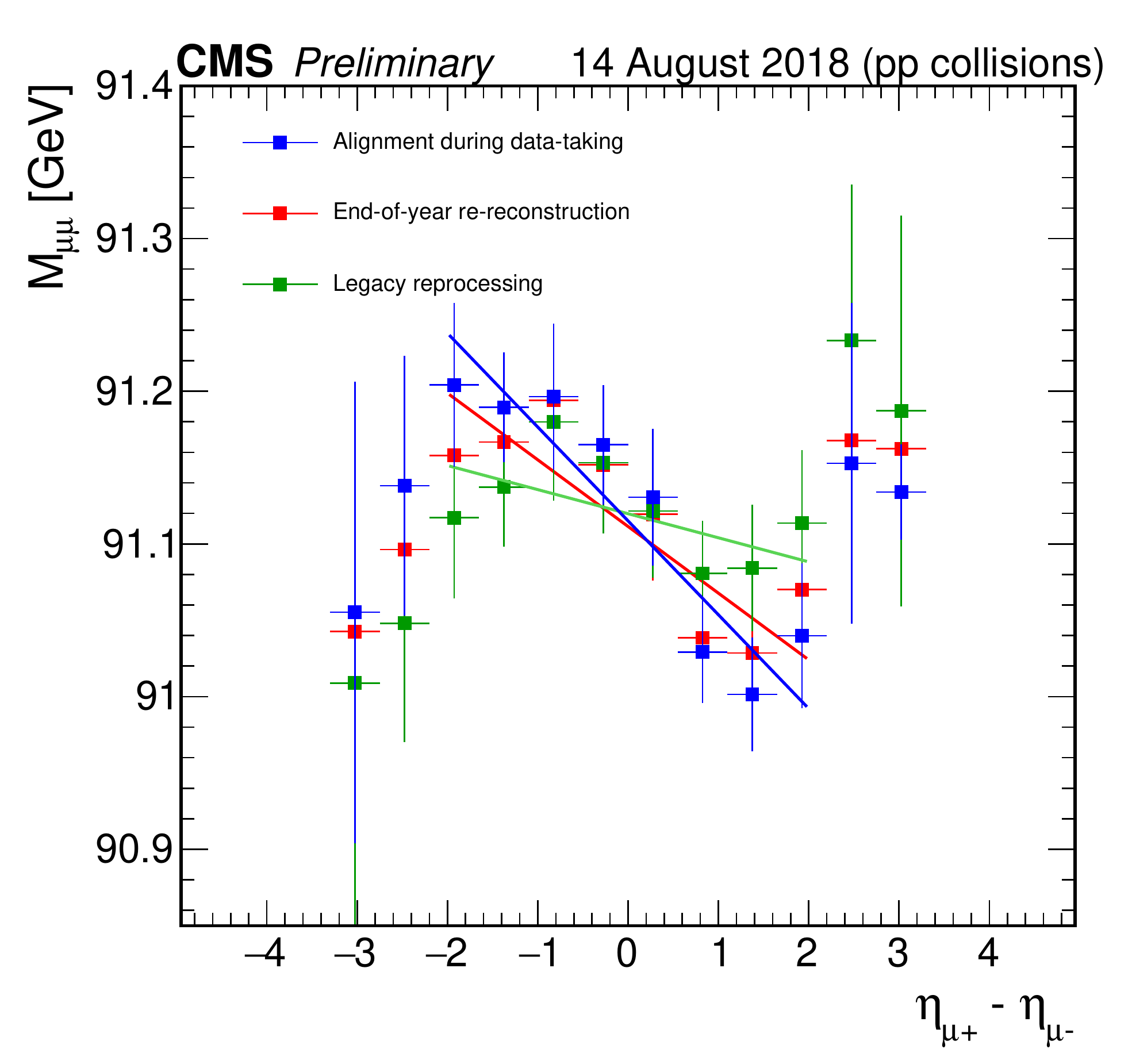}
       \caption{Reconstructed Z $\rightarrow \mu\mu$ mass as a function of the difference in pseudorapidity $\eta$ between positively and negatively charged muons.   The full sample of dimuon events in the 2016-2018 period was used \cite{citation5}.}
       \label{fig3}
     \end{figure}

\begin{figure}[!ht]
        \centering
      \includegraphics[width=0.8\textwidth]{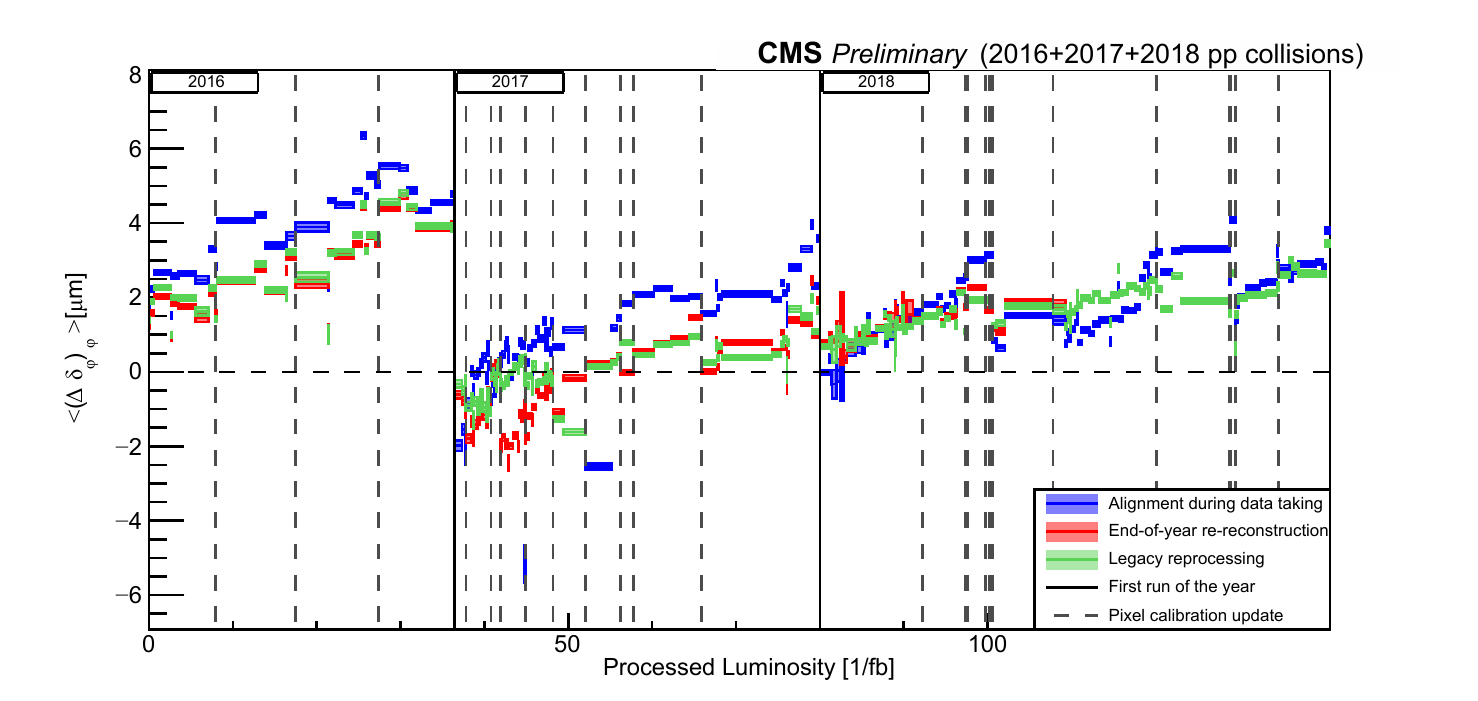}
      \caption{Mean difference in $\phi$ residuals for modules overlapping in the $\phi$ direction in the barrel pixel as a function of the processed luminosity \cite{citation5,citation6}.}
       \label{fig4}
     \end{figure}

Systematic distortions of the tracker geometry,  known as weak modes,  can cause biases in the track reconstruction,  affecting the performance of physics observables.  In an ideally aligned tracker, the reconstructed invariant mass of resonances such as the Z $\rightarrow\mu\mu$ decay should minimally depend on the azimuthal angle or the pseudorapidity of the muons. Therefore,  the reconstructed mass of the resonance serves as a good observable to inspect the geometry and detect the presence of weak modes.   The improvement for the legacy reprocessing in the uniformity of the reconstructed mass of the Z $ \rightarrow\mu\mu$ decay is demonstrated in Figure  \ref{fig3}.  The overlap validation constitutes another tool to monitor the alignment looking for systematic distortions, in this case by using hits from tracks that have two hits in separate modules within the same layer of the tracker.   Due to the small distance between the two hits,  there is a small uncertainty in the propagation of the track parameters, and the double difference of the track hit-residuals is very sensitive to systematic distortions.  The mean overlap residuals are shown in Figure \ref{fig4} as a function of the processed luminosity.  The overall reduction of the residual values can be observed for the legacy reprocessing.

\section{Run 3 prospects}
The integrated luminosity collected by the CMS experiment is expected to be doubled during Run 3 with respect to Run 2.  Stronger variations of the Lorentz drift of charged carriers released by charged particles passing through the silicon sensors are foreseen due to larger irradiation doses.  The alignment procedure is sensitive to Lorentz drift changes induced by accumulated radiation after $\approx$ 1 fb$^{-1}$,  while the pixel local reconstruction calibration which corrects for this effect is only performed after $\approx$ 10 fb$^{-1}$.  If the alignment is performed at a high enough granularity,  inward and outward-pointing modules are free to move separately and the bias coming from Lorentz angle miscalibration can be absorbed.  Therefore,  a finer granularity for the automated alignment is expected to be deployed during Run 3.  A finer granularity will be used at earlier stages with respect to Run 2 for the alignment run offline as well,  to better cope with radiation effects. In addition,  more geometries will be derived to cover the significant changes over time.

\section{Summary}
The performance of the tracker alignment corresponding to the final set of alignment conditions used for the legacy reprocessing of the CMS Run 2 data has been presented. The data-driven methods used to derive the alignment parameters and the set of validations that monitor the tracking and vertexing performance, as well as the presence of systematic distortions in the detector geometry, were reviewed.  Finally,  the prospects for the alignment calibration during Run 3, in particular, the challenges related to the radiation damage and the rapidly changing conditions were discussed.

\end{document}